\documentclass{iopart}
\usepackage{iopams}  
\usepackage{epsfig}

\begin{document}

\title[EAW physics in LPI]{Aspects of Electron Acoustic Wave Physics in Laser Backscatter from Plasmas} 
\author{N J Sircombe$^{1}$, T D Arber$^{1}$ and R O Dendy$^{2, 1}$}
\address{$^{1}$ Centre for Fusion, Space and Astrophysics, Department of Physics, University of Warwick, Coventry CV4 7AL, UK}
\address{$^{2}$ UKAEA Culham Division, Culham Science Centre, Abingdon, Oxfordshire, OX14 3DB, UK}
\eads{n.j.sircombe@warwick.ac.uk}

\begin{abstract}
Recent experimental results from the Trident laser confirm the importance
of kinetic effects in determining laser reflectivities at high intensities.
Examples observed include scattering from low frequency electron acoustic waves
(EAWs), and the first few stages of a cascade towards turbulence through
the Langmuir decay instability. Interpretive and predictive computational capability in this area
is assisted by the development of Vlasov codes, which offer high velocity
space resolution in high energy regions of particle phase space, and do not
require analytical pre-processing of the fundamental equations. A direct
Vlasov solver, capable of resolving these
kinetic processes, is used here to address fundamental aspects of the existence and stability of 
the electron acoustic wave, together with its collective scattering properties.
These simulations are extended to realistic laser and plasma parameters characteristic of  single hot-spot experiments. Results are in qualitative agreement with experiments displaying both stimulated Raman and stimulated electron acoustic scattering. The amplitude of simulated EAWs is greater than that observed experimentally, and is accompanied by a higher phase velocity. These minor differences can be attributed to the limitations of a one-dimensional collisionless model.\\
\end{abstract}



\maketitle

\section{Introduction}
Recent single hot-spot experiments using the Trident laser facility \cite{monty:2001, monty:2002} identified backscatter which resembles stimulated Raman scattering (SRS) but, importantly, is due to stimulated scattering from an electron plasma mode whose frequency is significantly below the plasma frequency. This mode was identified as the electron acoustic wave (EAW), an undamped electron mode absent from fluid descriptions of the plasma. In order to survive and propagate, the EAW requires a non-Maxwellian electron velocity distribution, flattened at the phase velocity of the wave, without which it would be critically damped. We demonstrate that in laser-plasma conditions which prohibit SRS, an incident electromagnetic wave can drive an initially critically damped electron plasma wave to sufficient amplitude that it can trap electrons, which then support it and allow it to propagate undamped, forming an EAW. Alternatively, flattened distributions supporting an EAW may be created by the trapping of electrons during the saturation of laser-plasma instabilities, such as conventional SRS in the case where plasma conditions permit it. 
The physical and mathematical characteristics of the EAW are described here, and the mode is then simulated using an Eulerian Vlasov code to demonstrate its undamped nature. The implications for laser plasma interactions are examined, including the Langmuir decay instability (LDI) and stimulated electron acoustic scattering (SEAS).

In particular, we model a scenario where the Langmuir wave excited by stimulated Raman scattering grows to an amplitude sufficient to cause a significant local flattening of the electron velocity distribution function. This local flattening is formally equivalent, as we shall see, to the creation of a small drifting beam population, and the EAW can be considered to be an electrostatic mode supported primarily by this beam population. Stimulated scattering of the incident laser light from the EAW then becomes possible (SEAS), by analogy with SRS. We show that numerical implementation of a model combining a kinetic (Vlasov) description of longitudinal dynamics with a fluid description of transverse dynamics successfully captures the key physics. 

In this paper we explore questions which, in addition to assisting the interpretation of present and possible future experiments in laser-plasma interactions, raise several interesting theoretical considerations. First, the literature on the theory and modelling of the EAW, which displays paradoxical qualities, is sparse, and for this reason we offer an \textit{ab initio} treatment. Second, SEAS provides interesting points of contact between the phenomenology of energetic particle populations in laser-plasma interactions and in magnetically confined fusion plasmas. These points of contact include: the existence of modes supported primarily by the energetic particle population; the independent role of such modes in coupling the plasma to external drivers; their role as a channel for energy transfer within the plasma, for example through a turbulent (or at least nonlinear) cascade; and their diagnostic potential. See, for example Refs. \cite{dendy:85,vann:03,vann:05,young:05} and references therein. Third, there is the question of what level of theoretical description (kinetic, fluid, etc.) best captures the key physics while also enabling effective numerical simulation.
\section{The Electron Acoustic Wave}
\label{TheEAW}
The possible existence of plasma waves at frequencies significantly below the electron plasma frequency was first identified by Stix\cite{stix}, although it was expected that Landau damping in this regime would prohibit their formation. Later work\cite{holloway:1989, holloway:1991, schamel:1986, schamel:2000} showed that EAWs can indeed exist, supported by a population of trapped electrons.
Derivation of the plasma dispersion relations, based on a two-fluid treatment, yields high frequency Langmuir waves and low frequency ion acoustic waves. A more complete kinetic treatment, based on linearising the Vlasov equation, shows these modes to be damped. This is due to Landau damping, a purely kinetic effect which (in one dimension for simplicity) requires that $\partial_v f|_{v_p} < 0$, where $f$ is the particle distribution function and $v_p$ the phase velocity of the wave. However if  $\partial_v f|_{v_p} = 0$ then the wave may be undamped and, as we shall see, this is the case for the EAW.

\begin{figure}
		\noindent
		\begin{center}
			\includegraphics[width=0.8\textwidth]{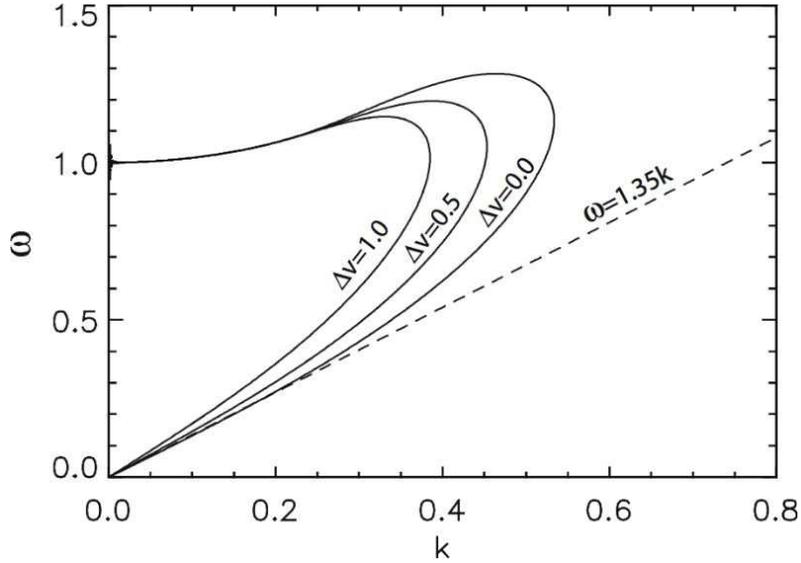}
			\caption{Dispersion relations calculated from Eq.(\ref{disp2}) for undamped plasma modes in the linear limit (where the width $\Delta v$ of the flattened region, in units of thermal velocity, tends to zero) and for two nonlinear cases ($\Delta v=0.5, 1.0$ in Eq.(\ref{f1})). Wavenumbers are normalised to $\lambda_D^{-1}$ and frequencies to $\omega_{pe}$. The  lower branch represents the electron acoustic wave, which for low $k$ follows $\omega = 1.35 k$. The upper branch represents an undamped form of the Langmuir mode.}
			\label{Fig1}
		\end{center}
	\end{figure}

In order to construct a dispersion relation for the EAW let us first consider a Maxwellian distribution with characteristic velocity $v_T$ that has a flattened region at $v=v_p$; we first take the limit where the width of the flattened region, in velocity space, tends to zero while $\partial_v f|_{v_p} = 0$. The integral in the Landau dispersion relation
\begin{equation}
\label{disp1}
\epsilon(\omega,k) = 1 - \frac{1}{2k^2\lambda_d^2}\int_{-\infty}^{\infty}\frac{\partial_v f(v)}{v-\omega/(k v_T)}dv
\end{equation}
can be written 
\begin{equation}
\label{disp2}
\int_{-\infty}^{\infty}\frac{\partial_v f(v)}{v-\omega/(k v_T)}dv
= P\int_{-\infty}^{\infty}\frac{\partial_v f(v)}{v-\omega/(k v_T)}dv + i\pi\partial_v f|_{v_p}
\end{equation}
By construction, the second term is zero, leaving only the principal value integral. Evaluating this integral gives the dispersion relation for EAWs in the linear limit:
\begin{equation}
\label{eaw}
k^2\lambda_d^2 + 1 - \sqrt{2}\frac{\omega}{k}\mathrm{Daw}\left(\frac{\omega}{\sqrt{2}k}\right)=0
\end{equation}
\begin{equation}
\label{daw}
\mathrm{Daw}\left(t\right)=\exp \left({-t^2}\right)\int_0^t\exp(u^2)du
\end{equation}
where Eq.(\ref{daw}) is Dawson's integral \cite{abramowitz}. Equation (\ref{eaw}), which can be evaluated numerically, thus gives the dispersion relation for undamped plasma waves in the limit of vanishing amplitude. The dispersion curve, labelled $\Delta v = 0$, is shown in Fig. \ref{Fig1}, where there are two distinct branches. The upper branch corresponds to an undamped form of the Langmuir wave and the lower branch, with $\omega<\omega_{pe}$, corresponds to the electron acoustic wave.

Similar analysis can be carried out for distribution functions that have a finite flattened region of width $\Delta v$ proportional to the wave amplitude, defined by
\begin{equation}
\label{flatf}
f = f_0 + f_1
\end{equation}
where $f_0$ represents a Maxwellian distribution, and flattening at $v=v_p$ is provided by
\begin{equation}
\label{f1}
f_1(v) = \partial_v f_0 |_{v_p}(v-v_p)\exp{\left(\frac{-(v-v_p)^2}{\Delta v^2}\right)}.
\end{equation}
Here the width of the flattened region is related to the EAW amplitude by assuming that electrons trapped in the wave potential can be considered as simple harmonic oscillators. Equating the potential and kinetic energies of a trapped electron then gives the width of the structure in velocity space, which dictates $\Delta v$ in Eq.(\ref{f1}). Equivalently one may calculate the maximum amplitude of an EAW which can be supported by a given flattened velocity distribution. 
We find, given Eq.(\ref{f1}), that $\Delta v = \sqrt{e\phi / m_e}$ where $\phi$ is the wave potential. Assuming a sinusoidal waveform this gives $\Delta v =  \sqrt{e E_0 / k m_e}$, where $E_0$ is the wave amplitude; we note that $k\Delta v$ equates to the nonlinear bounce frequency of resonant particles, as defined by Eq.(27) of Ref.\cite{berk:1995}, for example.
With $f_1$ defined in this way, Eq.(\ref{disp2}) can be solved numerically to give a family of dispersion curves inside the ideal, infinitesimal amplitude case, as shown in Fig. \ref{Fig1}. The dispersion relation of the EAW is thus linked to the wave amplitude.

\begin{figure}
		\noindent
		\begin{center}
			\includegraphics[width=0.8\textwidth]{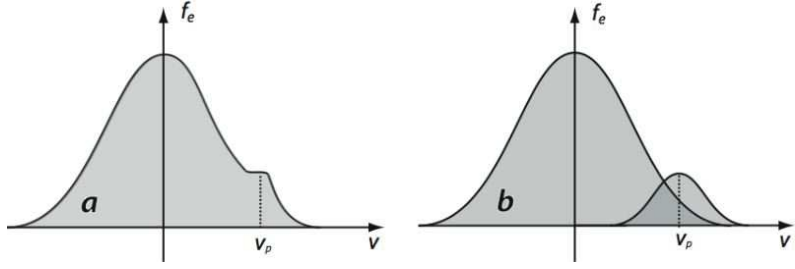}
			\caption{The distribution (a) which exhibits local flattening at $v_p$, is mathematically identical to (b), which comprises a background Maxwellian plus a smaller drifting electron distribution centred at $v_p$. The Doppler shifted frequency of plasma oscillations supported by the drifting electron distribution governs the EAW dispersion relation in the limit of small density.}
			\label{Fig2}
		\end{center}
	\end{figure}

The EAW appears, at first, to be unphysical. In particular its low frequency is a characteristic not expected of electron plasma waves supported by distribution functions arbitrarily close to Maxwellian, in which ion dynamics play no role. Some physical understanding, for the case of small $k$, can be gained by considering a locally flattened distribution to be a superposition of a background Maxwellian population and a smaller drifting population, as shown in Fig.\ref{Fig2}.

In the frame of reference of the background population, oscillations at the electron plasma frequency supported by the drifting population will be Doppler shifted such that $\omega = \omega'+kv$ where $\omega' = \left(e^2n_2 / \epsilon_0 m_e\right)^{1/2}$ is the electron plasma frequency for this second population, whose density is $n_2$. In the limit where $n_2$ tends to zero, we recover $\omega \sim vk$.
This dispersion relation is linear in {{\em k}} for small $k \lesssim 0.2\lambda_D^{-1}$, and implies a frequency below the plasma frequency. While this interpretation does not give the exact value for the phase velocity of  the EAW in the linear limit ($v_p\approx1.35 v_{Te}$, as in Fig.\ref{Fig1}),  it is helpful in understanding the origin of the low frequency modes described by Eq.(\ref{eaw}). This description also raises the question of negative energy modes, which can be supported in similar beam-plasma systems \cite{dendy}. Indeed, trapped electrons modes similar to the EAW described here can exist in a negative energy configuration \cite{grissmeier:2002}, although not in the regime considered here.

\section{Simulating an EAW}
\label{SimEAW}
\begin{figure}
		\noindent
		\begin{center}
			\includegraphics[width=0.8\textwidth]{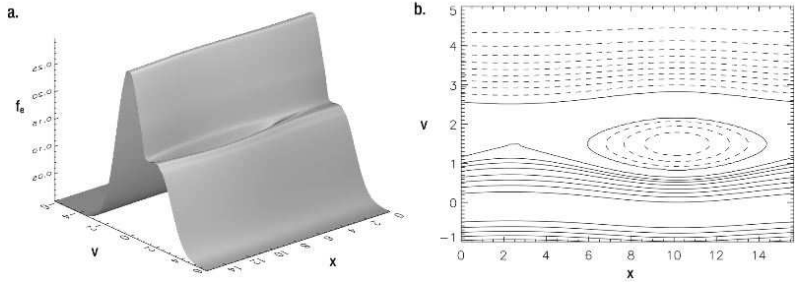}
			\caption{(a) Surface plot of the electron distribution function for a large amplitude EAW ($\delta n=0.1n_e, \omega =0.6 \omega_{pe}, k=0.4\lambda_D^{-1}$) at time $t=10^3\omega_{pe}^{-1}$, simulated using the Vlasov-Possion code. (b) Corresponding contour plot. Contours for $f_e < 0.135$ are drawn with dashed lines to highlight the trapped electron phase space structure. The $v$ axes are given in units of $v_{Te}$ and the $x$ axes in units of $\lambda_D$. System boundaries are periodic.}
			\label{Fig3}
		\end{center}
	\end{figure}

\begin{figure}
		\noindent
		\begin{center}
			\includegraphics[width=0.64\textwidth]{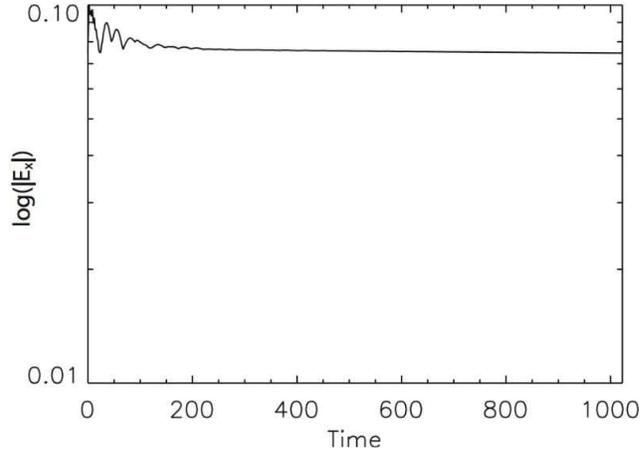}
			\caption{Logarithmic electric field amplitude of EAW  ($\delta n=0.1n_e, \omega =0.6 \omega_{pe}, k=0.4/\lambda_D$) against
			time. After an initial transient stage, the EAW persists as an electron plasma wave with frequency below the plasma frequency, which is undamped except for limited damping due to
			numerical diffusion.}
			\label{Fig4}
		\end{center}
	\end{figure}

A full treatment of the EAW requires a kinetic description of the plasma. This section outlines the fully kinetic Vlasov-Poisson model and the development of initial conditions for, and the simulation of, a travelling EAW.
The model used is a one-dimensional Vlasov-Poisson system of electrons and immobile protons with no magnetic field\cite{arber:2002}, in a numerical implementation which has been used previously to explore kinetic phenomena relevant to laser-plasma interactions\cite{sircombe:2005}. This fully nonlinear self-consistent system is governed by the Vlasov equation for the electron distribution function $f_e$,
\begin{equation}
\label{vlasov}
\frac{\partial f_e}{\partial t} + v\frac{\partial f_e}{\partial x} -\frac{e}{m_e}E\frac{\partial f_e}{\partial v} = 0
\end{equation}
and Poisson's equation for the electric field
\begin{equation}
\label{poisson}
\frac{\partial E}{\partial x} = -\frac{e}{\epsilon_0}\left(\int f_e dv - n_i\right)
\end{equation}
As an initial condition, let us consider an unperturbed distribution function flattened at a phase velocity $v_p$, such that $v_p = \omega / k$ where $\omega$ and $k$ can be chosen from the EAW branch of the dispersion relation. We specify
\begin{equation}
\label{fu}
f_u = f_0 + f_1
\end{equation}
where $f_0$ represents a Maxwellian distribution and $f_1$ is given by Eq.(\ref{f1}).
The width of the flattened region, which relates to the number density of trapped electrons, is given by $\Delta v$ and is proportional to the EAW amplitude (explicitly, $\Delta v =  \left({e E_0 / k m_e}\right)^{1/2}$). In order to create a travelling wave we perturb the distribution function given in Eq.(\ref{fu}) by setting $f_e = f_u + f_p$ where 
\begin{equation}
\label{fp}
f_p(x,v) = \frac{-eE_0}{m_e(\omega - kv)}\sin(kx)\partial_v f_u
\end{equation}
Equation (\ref{fp}) represents a finite-amplitude generalisation of a linear perturbation of the Vlasov equation about $f_u$. Note that Eq.(\ref{fp}) contains no singularities since $\partial_v f_u|_{v_p} = 0$ and $\omega$ is chosen to be real.
The Vlasov-Poisson system is initialised against a neutralising ion background in a periodic box with the distribution function $f_e = f_u + f_p$ and $\omega =0.6\omega_{pe}$, $k = 0.4\lambda_D^{-1}$, $\delta n = 0.1n_e$ where $\delta n = E_0 / L_x$. In this regime of $(\omega,k)$ we would not normally expect an electron plasma wave to propagate, undamped or otherwise. Figure \ref{Fig3} shows the trapped electron distribution of the EAW after a thousand inverse plasma angular frequencies. Figure \ref{Fig4} shows that the amplitude of the EAW is effectively constant: after an initial transient phase, only weak numerical damping remains. This numerical approach thus demonstrates how a non-Maxwellian distribution, specifically the flattening at the phase velocity of the wave introduced through Eqs.(\ref{f1}) and (\ref{fu}), is necessary for the propagation of an EAW. 

\section{Relevance to the Langmuir Decay Instability}
\begin{figure}
		\noindent
		\begin{center}
			\includegraphics[width=0.8\textwidth]{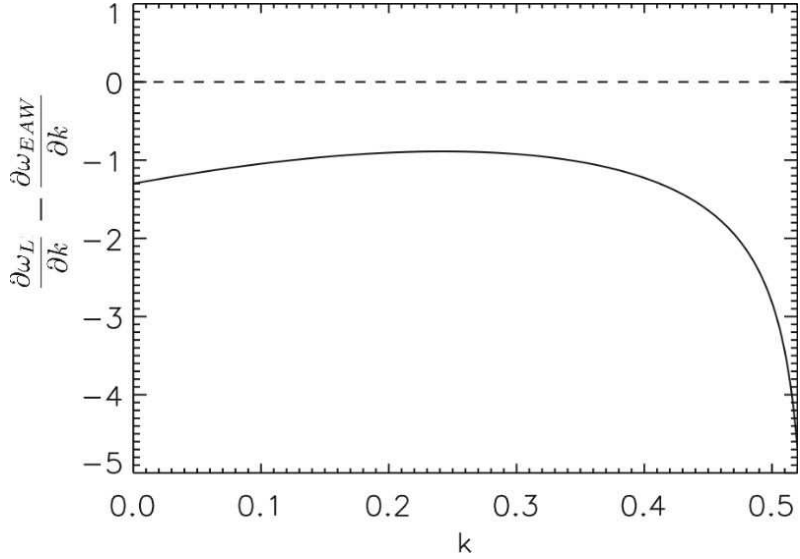}
			\caption{Difference between the gradients of the dispersion relations of the Langmuir and Electron Acoustic
			modes for a range of $k$. In order for Langmuir decay to occur, this quantity must be greater than zero. The
			critical wavenumber $k_c$ is the point at which the gradients are exactly equal. This curve remains negative
			for all k, demonstrating that straightforward Langmuir decay via the EAW, rather than the IAW, is not possible.}
			\label{Fig5}
		\end{center}
	\end{figure}

A Langmuir wave can decay into a second Langmuir wave of lower wavenumber plus an ion acoustic wave (IAW). This process can occur repeatedly, forming a Langmuir cascade \cite{depierreux:2000}. Can the EAW perform the role of IAW to produce a Langmuir cascade on electron timescales?
The conventional Langmuir cascade \cite{thornhill:1978} proceeds for all $k$ above a critical value $k_c$, determined by the point where the group velocity $\partial \omega / \partial k$ of the parent Langmuir wave (L) is equal to that of the IAW:
\begin{equation}
\label{kc}
\frac{\partial \omega_L}{\partial k} = \frac{\partial \omega_{IAW}}{\partial k} \Rightarrow k_c = \frac{1}{3\lambda_D}\sqrt{\frac{m_e}{m_i}}
\end{equation}
A similar analysis, for small $k$, can be performed in the case where the IAW is replaced by an EAW. Approximating the Langmuir dispersion relation by
\begin{equation}
\label{bgdisp}
\omega \approx \omega_{pe}\left(1+3k^2 \lambda_D^2 / 2\right)
\end{equation}
and the EAW dispersion relation by
\begin{equation}
\label{eawdisp}
\omega \approx \omega_{pe} \left(1.35k\lambda_D\right)
\end{equation}
gives a critical wavenumber $k_c \approx 0.45 \lambda_D^{-1}$, suggesting that LDI via the EAW might be a possibility. However, Eqs.(\ref{bgdisp}) and (\ref{eawdisp}) are no longer valid for such a high critical wavenumber. The assumption of small $k$ is therefore abandoned and the gradients calculated numerically to give Fig.\ref{Fig5}. It follows from this full treatment, valid for all $k$, that a process of Langmuir decay via the electron acoustic branch is not possible. However this does not rule out all forms of interplay between LDI and EAWs. The upper branch of the dispersion relation, essentially an undamped form of the conventional Langmuir mode, may replace one or both of the Langmuir waves in the LDI without affecting the critical wavenumber. This scenario is left for future work.

\section{Stimulated Electron Acoustic Scattering (SEAS)}
\label{SimSEAS}
The collective scattering of incident laser light from an EAW can be simulated using a Vlasov-Maxwell model. Here the relativistic Vlasov equation for electrons, in the presence of transverse fields
\begin{equation}
\label{emvlasov}
\frac{\partial f_e}{\partial t} + \frac{p_x}{m_e}\frac{\partial f_e}{\partial x} -\frac{e}{m_e}\left(E_x +v_y B_z\right)\frac{\partial f_e}{\partial p_x} = 0,
\end{equation}
is solved against a stationary ion background, together with Maxwell's equations
\begin{eqnarray}
\label{maxwell}
\frac{\partial E_y}{\partial t} & = & -c^2\frac{\partial B_z}{\partial x}-\frac{J_y}{\epsilon_0}\\
\frac{\partial B_z}{\partial t} & = & -\frac{\partial E_y}{\partial x}
\end{eqnarray}
in one dimension. Transverse motion of particles is treated as fluid-like, hence
\begin{eqnarray}
\label{fluid}
\frac{\partial v_y}{\partial t} & = & -\frac{e}{m_e}E_y\\
J_y & = & -en_e v_y
\end{eqnarray}
Poisson's equation (Eq.(\ref{poisson})) is solved as before, to give the longitudinal electric field.

The initial conditions are chosen to prohibit conventional stimulated Raman scattering (SRS), so that density is above quarter critical density, and to satisfy wavenumber and frequency matching conditions for SEAS. The system is periodic in $x$ and undriven, with a TEM wave present throughout the system initially. Vlasov codes are inherently noiseless, so a low amplitude density perturbation is added to a Maxwellian velocity distribution to seed the growth of the EAW. The wavenumbers and frequencies of the incident wave $E1$ ($k_{E1} = 0.815 \omega_{pe}c^{-1}$, $\omega_{E1} = 1.29 \omega_{pe}$), scattered wave $E2$ ($k_{E2} = -0.108 \omega_{pe}c^{-1}$, $\omega_{E2} = 1.0058 \omega_{pe}$) and the $EAW$ seed ($k_{EAW} = 0.923 \omega_{pe}c^{-1}$, $\omega_{EAW} = 0.2842 \omega_{pe}$) are chosen to satisfy the matching conditions for SEAS. The incident wave amplitude is $E_1 = 0.3 e/(\omega_{pe} c m_e)$ and the system length $L_x \approx 3084 c\omega_{pe}^{-1}$. The simulation is conducted on a numerical grid with 8,192 points in $x$ and 512 in $p$.

\begin{figure}
		\noindent
		\begin{center}
			\includegraphics[width=0.8\textwidth]{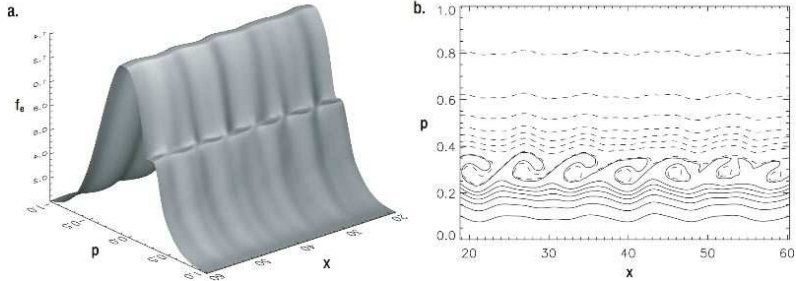}
			\caption{(a) Surface plot of the electron distribution function at $t=10^4\omega_{pe}^{-1}$. Only a small section of the complete system is shown for clarity. Electron trapping and flattening of the distribution function can be seen: this is the EAW which has grown from a background density perturbation as a result of SEAS. Axes are given in relativistic units, $c/\omega_{pe}$ for space and $m_e c$ for momentum. (b) Corresponding contour plot. Contours below $f_e=0.725$ are drawn with dashed lines to highlight the trapped electron holes. The $p$ axes are given in units of $m_e c$ and the $x$ axes in units of $c \omega_{pe}^{-1}$.}
			\label{Fig6}
		\end{center}
	\end{figure}

Figure \ref{Fig6} shows the electron distribution function at late time. The evolution of trapped electron structures, and resulting flattening of the distribution function, is visible, corresponding to an EAW.  We thus have SRS-like scattering in a plasma whose density is greater than quarter critical: this scattering is from an EAW, an electron plasma wave with a frequency below the plasma frequency.\\
This demonstrates the capability  of a Vlasov based kinetic plasma model to simulate the scattering of incident laser light from non-Maxwellian, trapped particle distributions. The development of such instabilities in realistic parameter regimes, and in the presence of an externally driven TEM wave, is now considered.
	
\section{Kinetic effects in single hot-spot experiments}
Non-Maxwellian particle distributions, which require a kinetic treatment of the plasma, can significantly affect the scattering of incident light, destroying the idealised picture of a three-wave parametric instability by allowing scattering from plasma oscillations omitted from conventional fluid treatments, such as the EAW. Section \ref{TheEAW} summarised the linear theory underpinning the EAW, section \ref{SimEAW} demonstrated its existence in the non-linear regime by way of electrostatic Vlasov-Poisson simulations, and section \ref{SimSEAS} confirmed the possibility of stimulated scattering, resembling SRS, from an EAW in a plasma of greater than quarter critical density. We now utilise an expanded Vlasov-Maxwell code to investigate SEAS and related kinetic effects in a regime close to that achieved in single hot-spot experiments \cite{monty:2001,monty:2002}. This involves less than quarter critical densities (hence permitting SRS) and the presence of a continuous EM driver.

\subsection{Numerical Approach}
The Vlasov-Maxwell code, described previously, is expanded to allow for the presence of a continuous, sinusoidal, EM driver at $x=0$. This requires that the system no longer be treated as periodic, instead the boundaries are open. Any charge flowing past the system boundaries is assumed then to reside on a \lq charged plate\rq, external to the system. This external charge is included when calculating the electrostatic potential in order to avoid the creation of a DC field. The electrostatic potential $\phi$ is found using a tridiagonal matrix inversion, and the electrostatic field is then given by $\partial_x \phi$. This replaces the Fourier method used to solve for $E_x$ in the periodic case, as described in Ref.\cite{arber:2002}.

\subsection{Initial conditions}
A system of normalised units is adopted in which time is normalised to units of $\omega_{pe}^{-1}$ and velocities to units of $c$. Thus space is normalised to units of $c\omega_{pe}^{-1}$, electric fields to units of $m_e c \omega_{pe} / e$ and magnetic fields to units of $m_e \omega_{pe} / e$. 
The laser intensity $I_0$, electron temperature $T_e$ and density $n_e$ achieved in single hot-spot experiments \cite{monty:2001, monty:2002} were, approximately:
\begin{eqnarray}
I_0 & = & 1.6\times 10^{16} \mathrm{Wcm^{-2}} \\
T_e & = & 350 \mathrm{eV}\\
n_e & = & 1.2\times 10^{20} \mathrm{cm^{-3}}=0.03n_c
\end{eqnarray}
These imply values for the simulation parameters (incident EM wave amplitude $E_{y_0}$ and frequency $\omega_0$, thermal velocity $v_{Te}$ and density $n_e$) of
\begin{eqnarray}
E_{y_0} & = & 0.33 m_e c \omega_{pe} / e\\
\label{om0}\omega_0 & = & 5.7775\omega_{pe} \\
v_{Te} & = & 0.026c\\
n_e & = & 1 \times \omega_{pe}^2 \epsilon_0 m_e / e^2 =0.03n_c.
\end{eqnarray}
To minimise the charge loss from the system, a \lq flat-top' density profile is used, where the density of both electrons and the neutralising ion background drops smoothly from $n_0$ to zero over a distance $\approx 40c\omega_{pe}^{-1}$  at the edges of the system. The simulation domain extends from $x=0$ to $x=220c\omega_{pe}^{-1}$, leaving a flat region at the centre of the simulation box approximately $140c\omega_{pe}^{-1}$ in length, from $p=-0.75m_ec$ to $p=0.75m_ec$. The simulation grid has $16,384$ points in $x$ and $1,024$ points in $p$. The simulation runs to an end time of $1200\omega_{pe}^{-1}$. 

\subsection{Results}
\begin{figure}
		\noindent
		\begin{center}
			\includegraphics[width=0.8\textwidth]{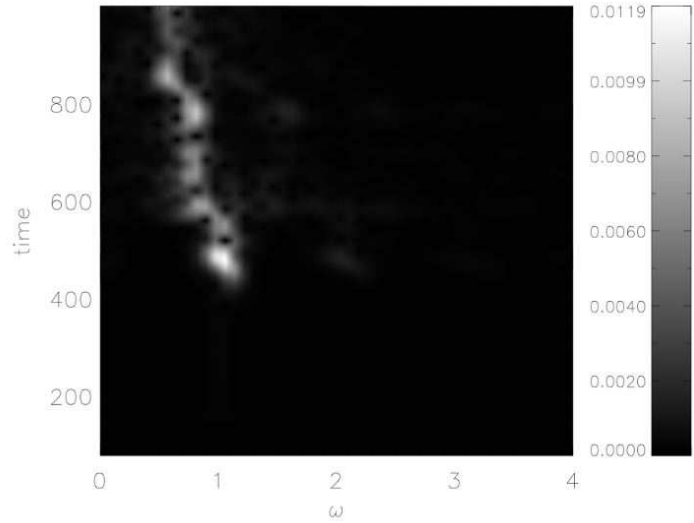}
			\caption{Windowed Fourier transform of the electrostatic field $E_x$ at the centre of the system. An initial SRS burst at $t \approx
			450/\omega_{pe}$ saturates via the trapping of electrons which distort the initially Maxwellian distribution
			and provide an environment in which waves below the plasma frequency can grow and propagate. The traces at
			$\omega \approx 0.8\omega_{pe}$ and $\omega \approx 0.6\omega_{pe}$, first appearing at
			$t\approx600/\omega_{pe}$, represent EAWs with phase velocities at $v_p = 2.73 v_{Te}$ and $2.03 v_{Te}$
			respectively.}
			\label{Fig7}
		\end{center}
	\end{figure}

\begin{figure}
		\noindent
		\begin{center}
			\includegraphics[width=0.8\textwidth]{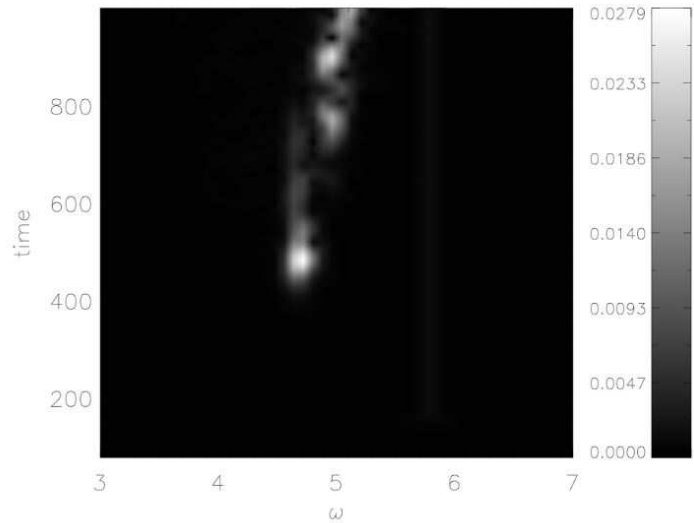}
			\caption{Windowed Fourier transform of the backwards propagating EM field at the centre of the system. The
			spectrum shows the light scattered by Langmuir waves (SRS) and EAW waves (SEAS) identified in the
			electrostatic spectrum at the same point in space (see Fig.\ref{Fig7}).}
			\label{Fig8}
		\end{center}
	\end{figure}

\begin{figure}
		\noindent
		\begin{center}
			\includegraphics[width=0.8\textwidth]{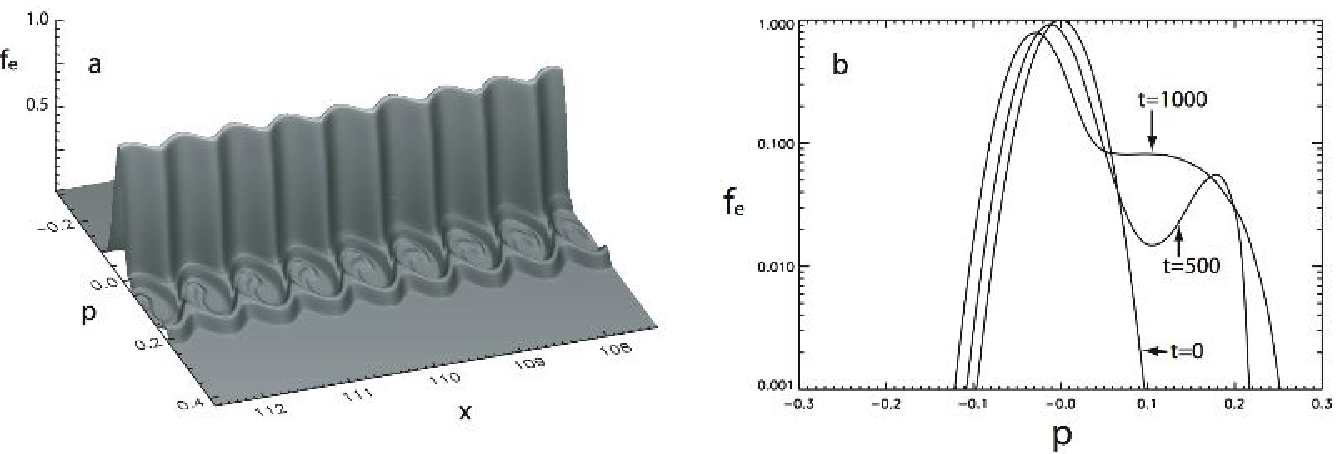}
			\caption{(a) Surface plot of the electron distribution near the centre of the system at $t=500/\omega_{pe}$. Electron trapping, visible here, is responsible for the saturation of the Raman instability and the creation of the electron beam in the spatially integrated distribution. (b) Spatially integrated electron distribution functions,  for $t=0$, $500\omega_{pe}$ and $1000\omega_{pe}$, normalised to the initial Maxwellian distribution. The trapping of electrons in the Langmuir wave driven by SRS temporarily creates a beam structure. The collapse of this structure is responsible in part for the formation of a broad plateau in momentum space at late times, which supports EAWs.}
			\label{Fig9}
		\end{center}
	\end{figure}

Figures  \ref{Fig7} and \ref{Fig8} display windowed Fourier transforms of the electrostatic field and of the back-propagating EM field, taken with a Hanning window of size $\approx 75\omega_{pe}^{-1}$, at the centre of the system. These show the development of low frequency plasma waves after $t=600\omega_{pe}^{-1}$.
In the initial SRS burst, starting at $t\approx 450\omega_{pe}^{-1}$ the EM driver at $\omega_0$ given by Eq.(\ref{om0}) scatters from a Langmuir wave at $\omega_1 = 1.06\omega_{pe}$, $k=0.27/\lambda_D$, $v_p = 3.93 v_{Te}$, to produce reflected light at a frequency $\omega_2 = 4.72\omega_{pe}$. This instability saturates via the trapping of electrons. Figure \ref{Fig9}a shows the electron distribution function during the late stages of the SRS burst, when electrons have been trapped and accelerated. A beam, similar to that observed in simulations of Raman forward scatter\cite{ghizzo:1990}, forms in the electron distribution which is clearly visible in plots (Fig.\ref{Fig9}b). The trapping of electrons by the Langmuir waves driven through SRS evolves into a plateau in the electron distibution. This flattened region extends to low phase velocities, providing an environment in which low frequency plasma modes are able to grow and propagate. These low frequency modes are visible in the electrostatic field spectrum after the collapse of the initial SRS burst at $t\approx 600\omega_{pe}^{-1}$, and correspond to two distinct electron acoustic waves (eaw1 and eaw2) at $\omega_{eaw1} = 0.73\omega_{pe}$, $k=0.27\lambda_D^{-1}$, $v_p = 2.73 v_{Te}$ and later $\omega_{eaw2} = 0.57\omega_{pe}$, $k=0.28\lambda_D^{-1}$, $v_p = 2.03 v_{Te}$.

The electron distribution at late times thus deviates significantly from a Maxwellian. The trapping of electrons in the initial SRS burst flattens the distribution around $p = 0.1$, allowing the development of low frequency plasma waves, the EAWs, whose trapped electrons further distort the distribution of particles. By the simulation's end, it has become clear that the plasma, and hence the modes which it supports, is not well described by linear or fluid approximations. Scattering observed in single hot-spot experiments was from EAWs with phase velocity $v=1.4v_{Te}$ ($k=0.29\lambda_D^{-1}$, $\omega = 0.41\omega_{pe}$), with a backscattered wave amplitude aproximately three thousand times smaller than that from SRS. The amplitude of EAWs, and of the light scattered from them, observed in simulations is greater than observed experimentally. The simulations presented here also produce EAWs with higher phase velocities (i.e. $v_p \approx 2.7 v_{Te}$ and $v_p \approx 2.0 v_{Te}$ compared to $v_p \approx 1.4 v_{Te}$) than the scattered spectra from experiments indicate. These two deviations are closely related. As shown earlier, the dispersion relation for the EAW is dictated in part by the mode amplitude. As the EAW amplitude is increased, the dispersion relation shifts inwards, as shown in Fig.\ref{Fig1}, resulting in a higher phase velocity at fixed wavenumber. Further work is required to quantify in greater depth this inconsistency between numerical and experimental results.
The simulation runtime, $t=1200\omega_{pe}^{-1}$, is equivalent to less than three picoseconds - this serves to highlight how rapid the switch from the fluid to the kinetic regime may be, at the laser intensities considered here. As laser intensity increases, the kinetic effects discussed here will become more critical to the understanding of the associated laser-plasma interaction physics.

\section{Conclusions}
Experiments studying fundamental laser-plasma interactions in a single hotspot \cite{monty:2001, monty:2002} observed backscattered light from the interaction of the incident beam with two distinct plasma modes. First, there is scattering from waves having high phase velocity $v_p \approx 4.2 v_{Te}$ and a frequency above the plasma frequency $\omega_{pe}$, which was attributed to SRS: the three-wave parametric instability involving a Langmuir wave. Second, there is scattering from waves of considerably lower phase velocity $v_p \approx 1.4 v_{Te}$, whose frequency is below $\omega_{pe}$, as low as $0.41\omega_{pe}$: these low frequency modes were identified as the electron acoustic wave. The simulations reported here have attempted to model the key physics of this scattering using a 1D Vlasov-Maxwell approach. These have been successful in achieving scattering from both high and low frequency electron plasma waves, but have not been able to reproduce exactly  the phase velocities of the EAWs and the relative amplitudes of the scattering events. 

The electron acoustic mode is a counter-intuitive phenomenon with a sparse literature: an electron plasma wave which propagates, free from Landau damping, at frequencies below the plasma frequency. This work has sought to clarify its characteristics, in terms of dispersion relations and the role of electron trapping, which also present an interesting application of plasma kinetic theory. Accurate representation and evolution of the complete phase space is of importance to SEAS, and is also vital to the saturation of the Raman scattering instability and the subsequent evolution of the system, as demonstrated here. Our simulations, in a regime close to those achieved in single hot-spot experiments, highlight the importance of kinetic effects, and the effect that the evolution of non-Maxwellian particle distributions may have on the scattering of incident light from an initially homogenous plasma. Even when the energy flow associated with SEAS is small, SEAS may have future applications as a diagnostic of the electron velocity distribution, given theoretical understanding and an appropriate modelling capability.

Recent work\cite{pesmesmall:2002,labaune:2004,glenzer:2004} has identified the need for a deeper understanding of laser-plasma interactions, particularly in the regimes currently being approached by the next generation of lasers. The accurate noise-free representation and evolution of the particle distribution functions provided by a Vlasov code make it a valuable additional tool complementing both fluid and particle-in-cell descriptions. While a full 3D Vlasov treatment is beyond the limits of current computing power, 1D and 2D Vlasov systems are tractable and can address many relevant problems. The present study has indicated some of the distinctive features of the EAW and SEAS physics that arise from the fact that the EAW can be considered to be primarily supported by an energetic particle population. 
Recent numerical work \cite{califano:2005} has highlighted the possibility that `trains' of electron holes, with low phase velocities, may be created by the action of a strong electrostatic driver at a frequency above the plasma frequency. Such structures could become involved in SEAS, and may be excited by the electrostatic daughter waves driven by the stimulated Raman and Brillouin instabilities. The interplay between the electrostatic mechanisms outlined in Ref.\cite{califano:2005} and the electromagnetic scattering mechanism outlined here is a potentially interesting topic for future work. Finally, some of the conceptual links to the role of energetic particle populations in magnetically confined plasmas were noted in the introduction, and these too may repay further investigation.

\ack
This work was supported in part by the Engineering and Physical
Sciences Research Council (EPSRC).

\section*{References}
\bibliographystyle{unsrt}

\end{document}